\documentclass[prb,showpacs,twocolumn]{revtex4}
\usepackage{epsfig}
\usepackage{graphicx}   
\usepackage{dcolumn}    
\usepackage{bm}         
\usepackage{color}
\usepackage{amssymb,amsmath}

\begin{document}
\title{Spatial ordering of nano-dislocation loops in ion-irradiated materials.\footnote{Corresponding author: S.L. Dudarev, email: sergei.dudarev@ccfe.ac.uk, tel. +44 1235 466513, fax. +44 1235 466435}}
\author{S.L. Dudarev$^1$, K. Arakawa$^2$, X. Yi$^{1,3}$, Z. Yao$^4$, M.L. Jenkins$^3$, M.R. Gilbert$^1$, and P.M. Derlet$^5$\\
$^1$EURATOM/CCFE Fusion Association, Culham Centre for Fusion Energy, Oxfordshire OX14 3DB, UK.\\
$^2$Department of Materials Science, Faculty of Science and Engineering, Shimane University, 1060 Nishikawatsu, Matsue 690-8504, Japan.\\
$^3$Department of Materials, University of Oxford, Parks Road, Oxford OX1 3PH, UK.\\
$^4$Department of Mechanical and Materials Engineering, Queen's University, Nicol Hall, 60 Union Street, Kingston K7L 3N6, Ontario, Canada. \\
$^5$Condensed Matter Theory Group, Paul Scherrer Institut, CH-5232 Villigen PSI, Switzerland.}

\begin{abstract}
Defect microstructures formed in ion-irradiated metals, for example iron or tungsten, often exhibit patterns of spatially ordered nano-scale dislocation loops. We show that such ordered dislocation loop structures may form spontaneously as a result of Brownian motion of loops, biased by the angular-dependent elastic interaction between the loops. Patterns of spatially ordered loops form once the local density of loops produced by ion irradiation exceeds a critical threshold value.
\end{abstract}


\maketitle

The use of ion irradiation experiments for simulating effects of neutron damage in materials has recently attracted renewed interest \cite{Kirk2009}, stimulated by practical advantages of the method, where a relatively high irradiation dose can be accumulated over only several hours of exposure to an ion beam, as opposed to several months or years of exposure typically associated with a neutron irradiation experiment. Ion irradiation has another advantage that the relatively low energy ions, as opposed to neutrons, do not initiate nuclear reactions and do not produce unstable radioactive isotopes during irradiation, hence not making the specimens radioactive.

The drawbacks associated with ion irradiation tests are (i) the small (micron) depth of penetration of ions into the material, requiring application of micro-mechanical methods for the investigation of mechanical properties of ion-irradiated samples, (ii) the proximity of surfaces, (iii) the fact that primary knock-on atom recoil spectra associated with ion impacts differ from recoil spectra generated by neutrons, and (iv) the high irradiation dose rate. The fact that dose rate significantly influences microstructure produced in {\it neutron}-irradiation materials tests, is well documented. Materials irradiated by neutrons at high dose rate show lower swelling than the same materials irradiated to the same integral dose at a lower dose rate \cite{Okita2002,Garner2004,Okita2007}. Irradiation embrittlement resulting from exposure to high-energy neutrons shows remarkable sensitivity to the dose rate, too, see for example Fig. 12 from Ref. \cite{Amayev1993}, where exposure to lower irradiation dose rate irradiation is shown to give rise to greater embrittlement. Similarly, low dose rate ion irradiation stimulates phase decomposition of Fe-Cr alloys\cite{Hardie2013}. The same alloys decompose less under high dose rate ion irradiation.

Given the highly non-linear nature of microstructural evolution under irradiation, the need to relate microstructure formed under ion irradiation to microstructure produced by neutron irradiation requires the development of computer models explaining the observed dose rate effects.
\begin{figure}[h]
\includegraphics[width=80mm]{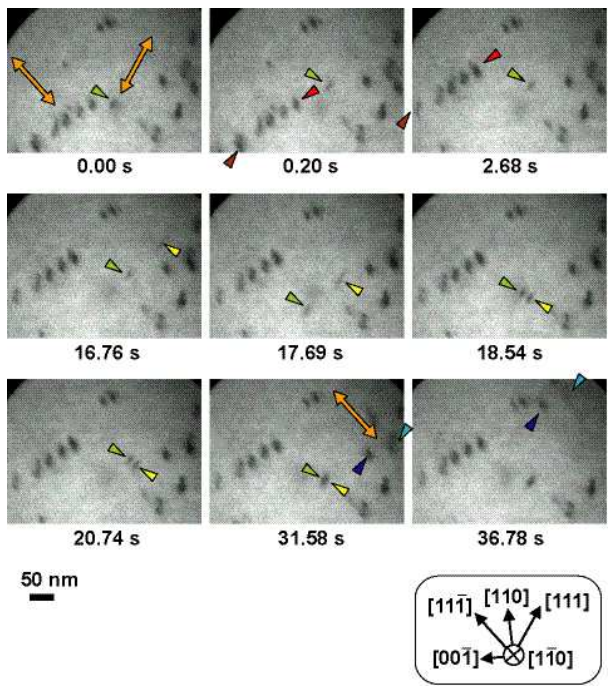}
\caption{\small A sequence of time frames taken using the ${\bf g}=110$ imaging condition at $T=725$K, which shows evolution of an ensemble of interacting mobile $a/2\langle 111\rangle$ prismatic dislocation loops in pure Fe. A bound configuration involving two loops is formed during the time interval between 16.76s and 18.54s. Coherent motion of a raft of loops is seen in the frames corresponding to the 0 - 2.68s time interval. An example of cooperative motion exhibited by a bound configuration of several loops is seen in the frames spanning the time interval between 31.58s and 36.78s.}
\label{Fig1}
\end{figure}

Within a rate theory treatment, involving single-particle distribution functions and assuming locally spatially homogeneous distribution of defect species, a higher defect production rate gives rise to the higher average concentration of radiation defects, which in turn increases the relative magnitude of non-linear terms in the rate equations, in particular the terms describing recombination of vacancies and self-interstitial atom defects, or the terms describing reactions between the defects \cite{Okita2002,Okita2007}. Elasticity effects are included in rate theory equations through mean-field bias factors that favour absorption of self-interstitial atoms defects by edge dislocations over absorption of vacancies \cite{Brailsford1981,Woo1981,Wolfer2007}.
What the rate theory analysis does not normally address is the evolution of a strongly spatially heterogeneous defect microstructure that forms as a result of elastic interaction between the defects, and the treatment of which requires the use of many-particle distribution functions, or an equivalent mathematical apparatus. Elastic interactions play a  particularly significant role in the case of dislocation loops that, as opposed to vacancies, interact even in an elastically isotropic material, for example in tungsten \cite{Eshelby1955,Hudson2005}.

\begin{figure}[h]
\includegraphics[width=80mm]{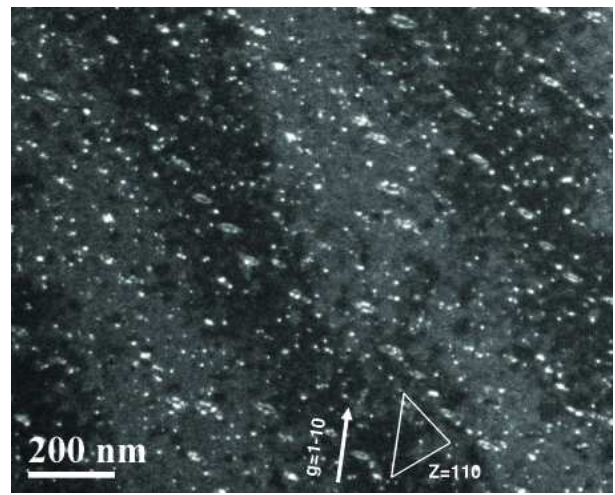}
\caption{\small Microstructure of ultra-high purity iron irradiated with Fe$^+$ ions to the dose of 1 dpa at 673K. The microstructure is formed by dislocation loops of interstitial type. The large loops seen in the micrograph have Burgers vectors of the $a/2\langle111\rangle$ type.}
\label{Fig2}
\end{figure}

\begin{figure}[h]
\includegraphics[width=80mm]{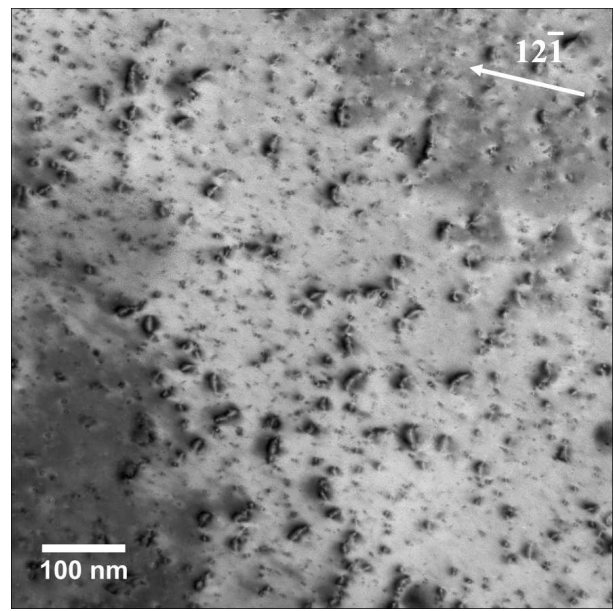}
\caption{\small Microstructure of 99.95 wt\% purity tungsten irradiated with 2 MeV W$^+$ ions to the dose of 3.3$\cdot 10^{17}$ W$^+$/m$^2$ at 773K. Incident ion beam direction is close to surface normal, which is parallel to the [113] crystallographic direction. Electron microscope observations were performed using the ${\bf g}=12\overline 1$ imaging condition. The loops that contribute to the formation of strings are of the $1/2\langle 111\rangle$ interstitial type.}
\label{Fig3}
\end{figure}

{\it In-situ} electron microscope examination of metals irradiated with ultra-high-energy electrons or energetic ions \cite{Arakawa2006,Arakawa2007,Arakawa2008,Yao2008,Hernandes2008,Jenkins2009,Yao2010,Arakawa2011,Yi2013} shows that nano-scale defects generated by irradiation perform stochastic Brownian motion (diffusion). If the density of defects is high, diffusion becomes biased by elastic interaction between the defects. The sequence of frames shown in Fig. \ref{Fig1}, and also electron microscope images of microstructure shown in Figs. \ref{Fig2} and \ref{Fig3}, illustrate how interaction between the defects affects microstructural evolution. The images show that in a dense ion-irradiation-induced microstructure nano-dislocation loops are not distributed evenly, and instead they form clusters, strings, and rafts \cite{Zinkle2006}. Furthermore, {\it in-situ} time-dependent electron microscope observations provide evidence that, depending on the relative position and orientation of their Burgers vectors, nano-scale dislocation loops interact through angular-dependent elastic forces, attracting or repelling each other.

Simulating real-time and real-space dynamics of microstructural evolution of an irradiated material still remains an outstanding problem. The ten orders of magnitude mismatch between the nanosecond timescale accessible to a conventional molecular dynamics (MD) simulation and the 10 to 1000 second timescale of a typical {\it in-situ} electron microscope observation often impedes meaningful quantitative analysis of experimental data. {\it In-situ} electron microscope observations show that dense microstructures formed under high irradiation dose rate conditions evolve through collective events, involving several defects moving in a correlated way \cite{Dudarev2010}, and resulting in the formation of rafts of nano-scale radiation defects.

While symmetry-broken solutions of reaction-diffusion rate theory equations were discussed in the past in connection with the treatment of irradiation-induced microstructures \cite{Ghoniem2002a,WooFrank1985,SemenovWoo2001,SemenovWoo2006a,SemenovWoo2006b}, the origin of spatial heterogeneity of rate theory solutions discussed in Refs. \cite{Ghoniem2002a,WooFrank1985,SemenovWoo2001,SemenovWoo2006a,SemenovWoo2006b} was associated either with the non-linearity of rate equations or with the anisotropy of diffusion of defects, and not with elastic interaction between the defects. {\it In-situ} observations illustrated in Fig. \ref{Fig1} suggest that it is the interaction between radiation defects that gives rise to the formation of locally spatially ordered defect structures. Interaction between radiation defects is expected to play a particularly significant role in the limit where the density of defects is high (this was noted already by Friedel \cite{Friedel1954} and Eshelby \cite{Eshelby1955}), i.e. in the limit of high irradiation dose rate.

In this paper, we investigate the stochastic dynamics of interacting nano-scale dislocation loops, extending and developing a method applied earlier\cite{Dudarev2010,Derlet2011} to the treatment of diffusion of nano-dislocation loops with collinear Burgers vectors, and exploring the three-dimensional dynamics of interacting loops with non-collinear Burgers vectors. In particular, we apply Langevin dynamics simulations to the interpretation of collective microstructural features found in {\it in-situ} electron microscope observations of irradiated materials illustrated in Figs. \ref{Fig1}, \ref{Fig2} and \ref{Fig3}.

Langevin dynamics simulations of interacting dislocation loops are based on solving coupled stochastic differential equations of the form
\begin{eqnarray}
{d r^{\alpha}_i\over dt}&=&-\sum _{\beta}{D^{\alpha \beta}_i \over k_BT}{\partial U\over \partial r^{\beta}_i}+\nu ^{\alpha}_i(t),
\label{Langevin}
\end{eqnarray}
where
\begin{eqnarray}
\langle \nu ^{\alpha}_i(t)\rangle &=&0,\nonumber \\
\langle \nu ^{\alpha}_i(t)\nu ^{\beta}_j(t')\rangle&=&2D_i^{\alpha \beta}\delta _{ij}\delta (t - t'),\label{correlator}
\end{eqnarray}
for $i=1,2,...,N$. The temperature-dependent diffusion matrix ${\bf D}_i=D_i^{\alpha \beta}=D_i^{\alpha \beta}(T)$ is related to the friction matrix ${\bm \gamma}_i=\gamma _i^{\alpha \beta}(T)$ (in dislocation dynamics ${\bm \gamma}$ is called the drag tensor) via the fluctuation-dissipation relation
\begin{equation}
{\bf D}_i (T)=k_BT [{\bm \gamma _i(T)}]^{-1}.
\end{equation}

\begin{figure}[t]
\includegraphics[width=80mm]{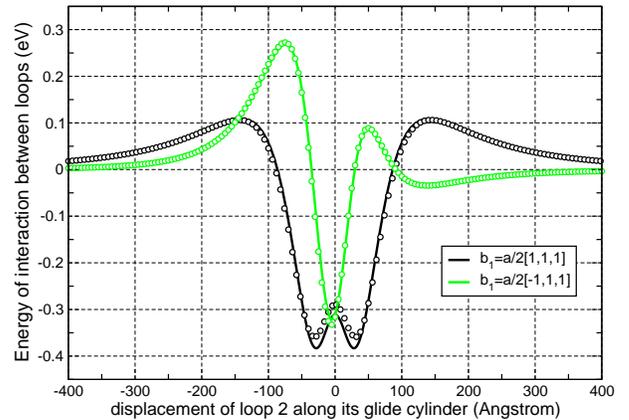}
\caption{\small Energy of interaction between two dislocations loops in Fe plotted as a function of the relative position of the loops in real space for the two orientations of the Burgers vector of the first loop, which is situated at the origin of the Cartesian system of coordinates ${\bf R}_1=0$. The energy of interaction, computed in the isotropic elasticity approximation, is shown as a function of the position of loop 2, which has Burgers vector ${\bf b}_2=a/2[111]$, along its glide cylinder. The glide direction of this loop intersects the $(x,y)$ plane at $(x,y)=(60,-60)$ \AA. The loop radii are $\rho_1=\rho_2=15$ \AA. Values of the shear modulus $\mu=82$ GPa (in atomic units $\mu=0.511$ eV/\AA$^3$) and the Poisson ratio $\sigma=0.29$ for iron are taken from Appendix 1 of Ref. \cite{HirthLothe}. Open circles represent interaction energies computed numerically using the double integral expression (\ref{U12}), lines correspond to a simplified analytical expression for the interaction energy.}
\label{Fig4}
\end{figure}

Elastic interaction between the loops is given by the following equation, derived using the isotropic elasticity approximation \cite{HirthLothe}
\begin{eqnarray}
U_{12}&=&-{\mu \over 2\pi}\oint \limits _{C_1}\oint \limits _{C_2}{({\bf b}_1\times{\bf b}_2)(d{\bf l}_1\times d{\bf l}_2) \over r}
\nonumber \\
&+&{\mu \over 4\pi (1-\sigma)}\oint \limits _{C_1}\oint \limits _{C_2}({\bf b}_1\times d{\bf l}_2)\overleftrightarrow T({\bf b}_2\times d{\bf l}_1).\label{U12}
\end{eqnarray}
We assume that the loop Burgers vectors are normal to the loops habit planes, i.e. $({\bf b}_1\cdot d{\bf l}_1)=0$,  $({\bf b}_2\cdot d{\bf l}_2)=0$. Integration in equation (\ref{U12}) is performed over closed contours $C_1$ and $C_2$,
corresponding to the perimeters of the loops. $\mu$ is the shear modulus and $\sigma$ is the Poisson ratio of the material.  Tensor $\overleftrightarrow T=T^{\alpha \beta}$ has the form
\begin{equation}
T^{\alpha\beta}={\partial ^2 r\over \partial r^{\alpha}\partial r^{\beta}}={r^2\delta^{\alpha \beta}
-r^{\alpha}r^{\beta}\over r^3}.
\end{equation}
One of the significant features characterizing the law of elastic interaction between prismatic loops (\ref{U12}) is its pronounced angular character, resulting in the energy of interaction depending not only on the orientation of the loop normal vectors, but also on the orientation of these vectors with respect to the vector ${\bf R}_1-{\bf R}_2$ linking the centers of the loops, as illustrated in Figure \ref{Fig4}. Fig. \ref{Fig4} shows that the energy of elastic interaction plotted as a function of position of the centre of one of the loops depends sensitively not only on the relative position of the loops but also on the orientation of their Burgers vectors.
Furthermore, the figure shows that this interaction between nano-dislocation loops varies on the electron-volt scale, which is not dissimilar to the scale of activation energies characterizing thermally activated migration of point defects, see e.g. Tables 1 and 2 of Ref. \cite{Dudarev2013}. Analysis performed in Ref. \cite{Dudarev2010} already showed that elastic interaction between mobile dislocation loops and vacancy clusters can trap mobile loops on the 100 second, or longer, timescales. A similar argument applies to self-trapping of interacting loops in the ``center of mass" frame, where two or several loops can form a bound configuration and migrate as one collective entity on the timescales of seconds, minutes, or even hours.

To understand the origin of this self-trapping, we estimate the characteristic confinement time for a loop trapped by one of the potential wells shown in Fig. \ref{Fig4}. The rate of escape from a potential well by diffusion can be estimated as
\begin{equation}
\Gamma \sim {D\over W^2}\exp(-E_b/k_BT),\label{escape}
\end{equation}
where $D$ is the loop diffusion coefficient, $W$ is the effective width of the potential well, and $E_b$ is the effective energy barrier that a loop needs to overcome to escape from the potential well. We assume that the width of the potential well is $W\approx20$ nm (this corresponds to the collinear orientation of Burgers vectors of the loops in Fig. \ref{Fig4}), and that the diffusion coefficient $D(T)$ is given by equation \cite{DofT}  $D(T)=(\overline D/2\rho)\exp(-E_a/k_BT)$, where $\overline D=5.7\cdot 10^{13}$ (nm)$^3$/s, $\rho$ is the loop radius (in nm units), $E_a=1.3$ eV is the activation energy for loop migration, and $E_b\approx 0.4$ eV  (this again corresponds to the collinear Burgers vectors of loops in Fig. \ref{Fig4}). Using equation (\ref{escape}), we find that the lifetime $\tau = 1/\Gamma$ of a self-trapped loop configuration at various temperatures is $\tau \sim 3$s for $T=500^{\circ}$C, $\tau \sim 130$s for $T=400^{\circ}$C, and $\tau \sim 2.2\cdot 10^4$s (or, in other words, $\tau \sim6.25$ hours) for $T=300^{\circ}$C. These estimated reaction-diffusion timescales are broadly similar with what is observed experimentally, see for example Fig. \ref{Fig1}, and other studies exploring the dynamics of migration of nano-dislocation loops in ion-irradiated iron \cite{Yao2008} and tungsten \cite{Yi2013}.

The estimates for elastic self-trapping timescales given above involve diffusion coefficients for migrating loops derived from experimental observations of Brownian motion of the loops \cite{Arakawa2007}. The activation energy $E_a=1.3$ eV  characterizing Brownian motion of loops seen experimentally is much higher than the activation energy $\sim 0.024$ eV found in molecular dynamics simulations of defects in pure iron \cite{Osetsky2003}. To find out whether impurities are responsible for the difference, we use an equation \cite{Dudarev2013} that relates the concentration of impurities $c$, in atom per site units, near a defect (for example a dislocation loop) to their concentration $c_0$ far away from a defect,
\begin{equation}
c={c_0\exp\left(-{U\over k_BT} \right) \over 1+c_0\left[\exp\left(-{U\over k_BT} \right) -1 \right]}.
\end{equation}
Here $U$ is the energy of interaction between an impurity and a defect and $T$ is the absolute temperature. For $c_0=10^{-6}$, $U=-0.66$ eV (the energy of interaction between an edge dislocation and a carbon atom in bcc iron \cite{Clouet2008}) and $T=500$ K, we find that $c\approx 0.82$. In other words, in thermodynamic equilibrium carbon impurities form a dense atmosphere around a dislocation loop even if the average concentration of carbon atoms in iron is as low as $c_0=10^{-6}=1$ appm. This shows that Langevin dynamics simulations, describing evolving defect configurations in the long time scale limit, should necessarily use dislocation mobilities that take into account the effect of impurity atmospheres.

Among various orientations of the loops' Burgers vectors, the collinear configurations exhibit the highest binding energy, and also the width of the potential wells describing elastic interaction between the loops is the largest in the collinear case. Noting this fact and relating it to equation (\ref{escape}), we arrive at a conclusion that an ensemble of the initially randomly distributed mobile interacting dislocation loops should exhibit a tendency towards local ordering, where loops with collinear Burgers vectors form collective relatively stable raft-like configurations. This hypothesis is confirmed by direct simulations, where we analyze solutions of Langevin dynamics equations (\ref{Langevin}) for a large ensemble of migrating nano-dislocation loops. We find that loops with collinear Burgers vectors do form bound configurations, which remain stable over fairly long intervals of time, whereas the loops with non-collinear Burgers vectors do not exhibit any pronounced tendency towards self-confinement and spatial ordering.

\begin{figure}[h]
\includegraphics[width=60mm]{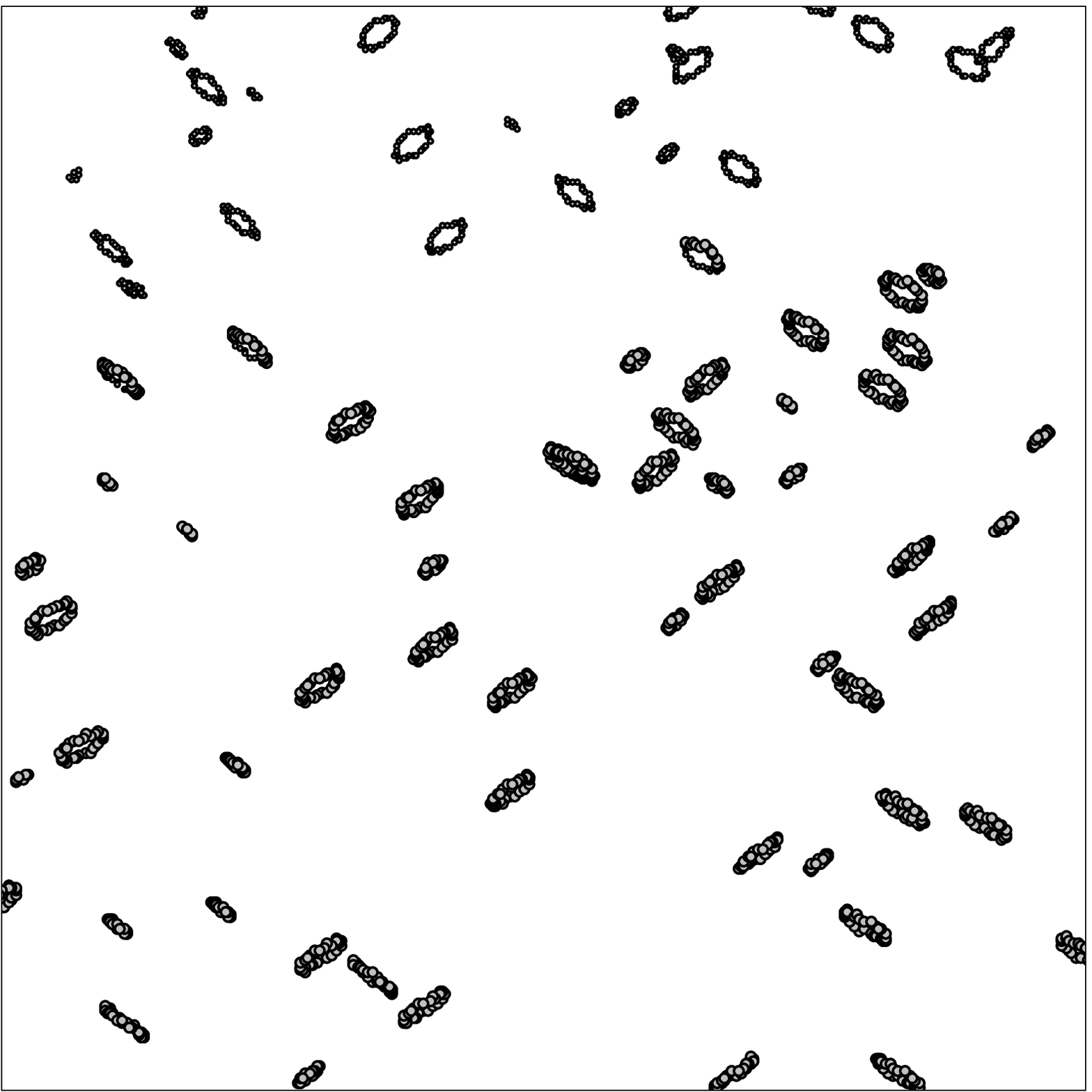}
\vskip 1cm
\includegraphics[width=60mm]{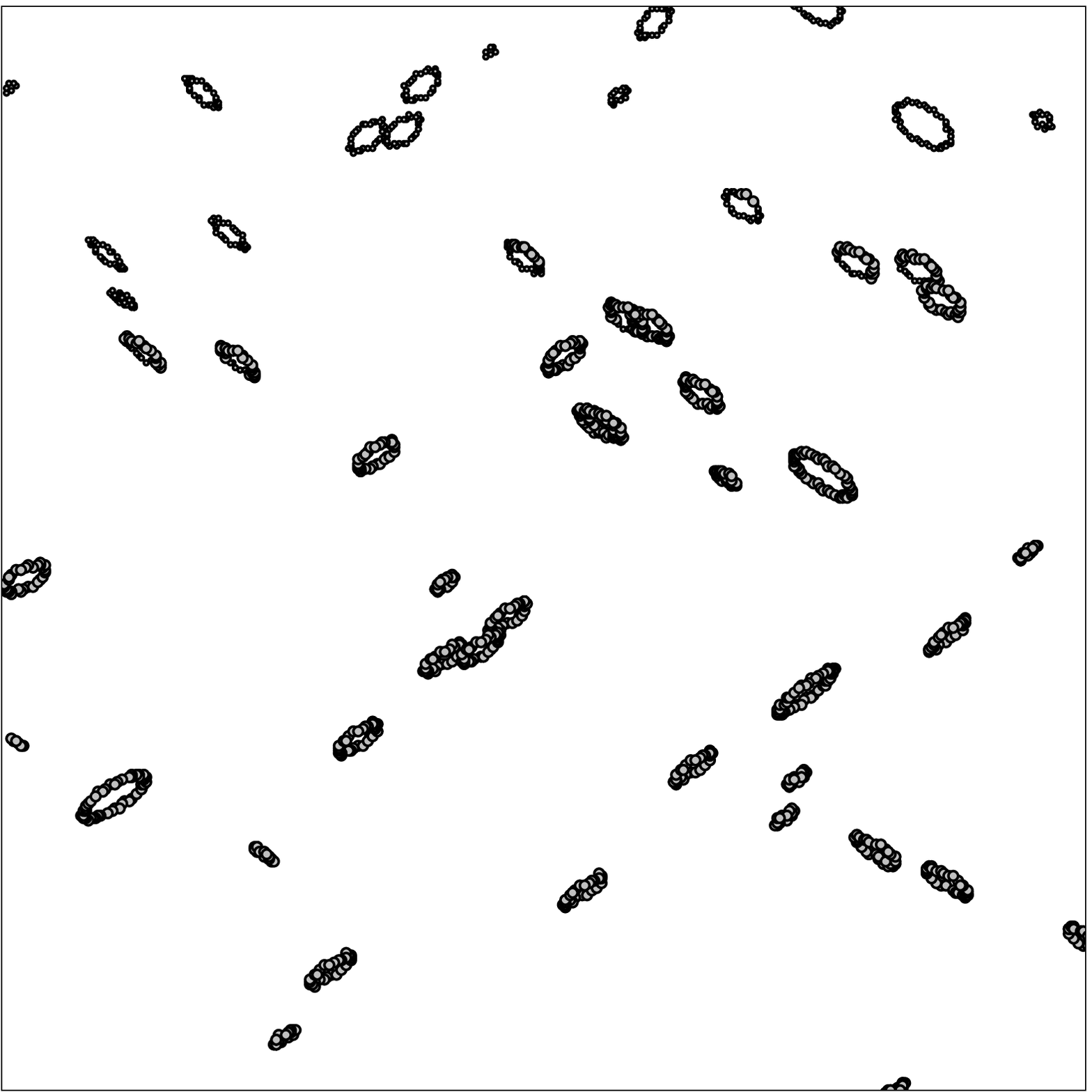}
\caption{\small Snapshots illustrating evolution of an ensemble of interacting nano-dislocation loops simulated using the Langevin dynamics equations (\ref{Langevin}). The loops are initially distributed randomly in a planar slab of Fe, treated as an elastically isotropic medium. The loops are confined between the two $(011)$-type planes, to mimic a foil bound by the $(011)$-type surfaces. No periodic boundary conditions are applied; the viewing direction is normal to the slab surface. The loop radii are chosen at random in the interval between 1 nm and 5 nm, and the loop Burgers vectors ${\bf b}_i$ are chosen randomly between $a/2[1\overline 11]$ and $a/2[11\overline 1]$. The loops interact according to equation (\ref{U12}), where the shear modulus $\mu=82\cdot10^9$ Pa (in atomic units $\mu=0.511$ eV/\AA$^3$) and the Poisson ratio $\sigma=0.29$ are taken from Appendix 1 of Ref. \cite{HirthLothe}. The temperature of the system is $T=670$K. The top panel shows the initial configuration of the system, the bottom panel represents a configuration formed after approximately 40s. Note the spontaneous formation of quasi-stable string structures formed by loops with collinear Burgers vectors.}
\label{LangevinEnsemble}
\end{figure}

Figure \ref{LangevinEnsemble} illustrates a typical behaviour of an ensemble of interacting nano-dislocation loops, evolving according to equations (\ref{Langevin}). The evolution of the system is conservative in the sense that the total number of self-interstitial atoms forming the loops (or, in other words, the total area of the loops) remains constant throughout the simulation. We see that the evolution of the system is continuous, and no stationary steady-state is reached on the timescale of the simulation. This is fundamentally consistent with the diffusive nature of evolution where, because of the thermal noise, there is no possibility of arriving at a steady-state configuration of the system. Simulations show the spontaneous formation of locally spatially ordered {\it many-body} collective loop structures, involving loops with collinear Burgers vectors, which remain stable over long intervals of time. These configurations form spontaneously in the simulations, similarly to how they form in {\it in-situ} observations, see Fig. \ref{Fig1}. The collective loop configurations eventually decay as a result of loops escaping by diffusion from the action of elastic forces holding them together. In agreement with the qualitative argument given in the preceding section, the 3D Langevin dynamics simulations show that such a self-organization effect, namely the formation of collective loop raft structures, only involves loops with collinear Burgers vectors. This is consistent with the analysis of the Burgers vector content of loop rafts observed experimentally.

The effect of formation of rafts, resulting from the combined action of elastic forces and Brownian motion of loops, is ultimately related to the fact that the density of loops is sufficiently high. Only in the limit of high density of defects, where the average distance between clusters of defects, for example nano-scale dislocation loops, is small, do elastic interactions play a significant part. This highlights the need to include the treatment of elastic interactions between radiation defects in the models for microstructures formed in ion irradiation experiments, where the high dose rate of irradiation produces high densities of defects. Also, the above analysis shows that interaction and reactions between radiation defects should be expected to drive microstructural evolution towards the formation of complex self-organized configurations and microstructural instabilities -- the investigation of which deserves greater effort, focused on unraveling the link between the microscopic laws of interaction between radiation defects and visible macroscopic manifestations of radiation damage.

\begin{acknowledgments}
The authors gratefully acknowledge discussions with A.A. Semenov, C.H. Woo, S.J. Zinkle, E. Clouet, F. Legoll, and N. Ghoniem.  This work was part-funded by the RCUK Energy Programme [grant number EP/I501045] and by the European Union's Horizon 2020 research and innovation programme.  To obtain further information on the data and models underlying this paper please contact
PublicationsManager@ccfe.ac.uk.  The views and opinions expressed herein do not necessarily reflect those of the European Commission. This work was also part-funded by the EPSRC via a programme grant EP/G050031. We acknowledge support from Grant-in-Aid for Scientific Research (Grants No. 24360395, 24656371, 24560805 and 22224012) from the Ministry of Education, Sports, Culture, Science and Technology in Japan, CREST, Japan Science and Technology Agency, and Nippon Steel \& Sumitomo Metal Co. Research Promotion Grant. We acknowledge support from Argonne National Laboratory (US) and the National Ion Beam Center (UK) where ion beam irradiation experiments were performed.

\end{acknowledgments}


\begin{thebibliography}{99}
\bibitem{DofT} Derived using an argument that the diffusion coefficient is inversely proportional to the friction coefficient, which is in turn proportional to the perimeter of the loop.
\end{thebibliography}
\end{document}